\documentclass[conference]{IEEEtran}
\IEEEoverridecommandlockouts
\usepackage{cite}
\usepackage{amsmath,amssymb,amsfonts}
\usepackage{algorithm,algorithmic}
\usepackage{graphicx}
\usepackage{xcolor}
\usepackage{hyperref}
\usepackage{caption}
\usepackage[belowskip=2pt,aboveskip=3pt]{caption}
\usepackage{textcomp}
\usepackage{amsmath}

\def\BibTeX{{\rm B\kern-.05em{\sc i\kern-.025em b}\kern-.08em
    T\kern-.1667em\lower.7ex\hbox{E}\kern-.125emX}}
\begin{document}
\setlength{\textfloatsep}{0pt plus 2pt minus 2pt}

%\title{Mitigating 5G Registration Signaling Storm using Blockchain}
\title{Who should I Collaborate with? \\ A Comparative Study of Academia and Industry Research Collaboration in NLP}

\makeatletter
\newcommand{\linebreakand}{%
  \end{@IEEEauthorhalign}
  \hfill\mbox{}\par
  \mbox{}\hfill\begin{@IEEEauthorhalign}
}
\makeatother

\author{\IEEEauthorblockN{Hussain Sadiq Abuwala, \; Bohan Zhang, \; Mushi Wang}
    \IEEEauthorblockA{\textit{Cheriton School of Computer Science} \\
    \textit{University of Waterloo}\\
    Waterloo, Canada \\
    \{hsabuwal, bohan.zhang, m358wang\}@uwaterloo.ca}
}

% \author{\IEEEauthorblockN{Bohan Zhang}
%     \IEEEauthorblockA{\textit{Cheriton School of Computer Science} \\
%     \textit{University of Waterloo}\\
%     Waterloo, Canada \\
%     bohan.zhang@uwaterloo.ca}
%     \and
% \IEEEauthorblockN{Paul Zeinaty}
% \IEEEauthorblockA{\textit{Department of Computer Science} \\
% \textit{Ecole Polytechnique, Institut Polytechnique de Paris}\\
% Palaiseau, France \\
% paul.zeinaty@polytechnique.edu}
% \and
% \IEEEauthorblockN{Noura Limam \;\;\; Raouf Boutaba}
% \IEEEauthorblockA{\textit{Cheriton School of Computer Science} \\
%     \textit{University of Waterloo}\\
%     Waterloo, Canada \\
%     \{noura.limam, rboutaba\}@uwaterloo.ca}
% }

\maketitle

\section{abstract}
The goal of our research was to investigate the effects of collaboration between academia and industry on Natural Language Processing (NLP). To do this, we created a pipeline to extract affiliations and citations from NLP papers and divided them into three categories: academia, industry, and hybrid (collaborations between academia and industry). Our empirical analysis found that there is a trend towards an increase in industry and academia-industry collaboration publications and that these types of publications tend to have a higher impact compared to those produced solely within academia.

\section{Introduction}
A typical mission statement from an academic institution can involve phrases like “commitment to excellent teaching”, “innovative research” and “personal and academic development of its students”. On the other side of the spectrum, delivering shareholder value is a firm's main goal. Start-ups are required to give their investors returns. There’s a dichotomy between the openness culture shared by academia, where they are encouraged to publish new findings, and more of a guarded culture shared in the corporate world, where their primary aim is to monetize inventions. Can they ever cooperate given their opposing goals, let alone work together?

There have been a few studies that have already looked into this and have found that collaboration between industry and academia has been on the rise. A study done in the Nature journal \cite{http1} found out that partnerships between academia and industry more than doubled between 2012 and 2016 in subjects such as Chemistry, Life Sciences, Physical Sciences and Earth \& Environmental sciences. This can be attributed to the fact that universities and businesses are both experiencing fast cultural change, and corporate cultures are increasingly embracing openness. Universities are also becoming aware of the need to find a means to make their research more socially relevant. 

In this paper, we focus on the field of Natural Language Processing(NLP). Although both academia and industry have different goals, their skill sets are complementary. Each contributes in some way to cutting-edge findings. Companies have more computing resources and private data sources. This allows them to train models faster and more efficiently and reduces the cost of trial and error. Moreover, from the market experience, they have a clearer vision, understand what kind of applications people need, and what problems are more worth solving, which is critical in the applied machine learning field. On the other hand, university researchers have no research restrictions and can freely explore the problems they are interested in. They have more research experience and are allowed to spend a long time working on one problem. It is also easier for them to find an expert in school to ask for help and collaboration. 

% With the goal to analyze collaboration patterns and citation rates between industry and academia. In particular, we introduce a novel dataset that comprises NLP papers (from ACL Anthology) divided into 3 distinct categories of collaboration: 1) Academia - papers that have co-authors only from academia, 2) Industry - papers that have co-authors only from industry, 3) Hybrid - papers that have co-authors from both academia and industry; and seek to compare the collaboration patterns and citation rate between the aforementioned categories. The dataset and the analyses have many uses, among which include understanding how the field of NLP is growing in terms of people working in their own domains (Academia or Industry) and people working across domains (Academia-Industry collaboration); and also quantifying the impact of papers published based on the 3 aforementioned collaboration patterns (Academia, Industry and Hybrid) using citations.

The goal of this paper is to analyze collaboration patterns and citation rates between industry and academia. In particular, our contributions are:

\begin{enumerate}

  \item   Introduced a pipeline to extract affiliations and citations from papers and classify them into three collaboration categories of Academia, Industry, and Hybrid (Academia-Industry collaboration).

  \item Empirically show that there is an increasing trend of Industry and Academia-Industry collaboration publications.

  \item Empirically show that Industry and Academia-Industry collaboration publications are overall higher in impact compared to Academia papers.

\end{enumerate}

\section{Related Work} 
A digital library of free-to-access articles on NLP in the public domain is called the ACL Anthology (AA). It covers papers presented at conferences in the ACL family and other NLP conferences like LREC and RANLP. Currently, 80957 papers on computational linguistics and natural language processing are available in AA, dating back to articles published since 1965. It is by far the most comprehensive single repository of scientific material on NLP. In the past, many AA subsets have been employed for various purposes. This includes summary generation of scientific articles \cite{qazvinian2013generating}, where citations were used to produce automatically generated technical extractive summaries, creating a corpus of scientific papers in \cite{mariani2019nlp4nlp}, the study of citation patterns and intent in \cite{radev2016bibliometric}, \cite{zhu2015measuring} and monitoring the ebb and flow of NLP topics, the interaction across subfields, and the impact of scholars from outside NLP \cite{anderson2012towards}. The work closest to us is in \cite{mohammad2020examining}, where the authors examined the trends in citations of NLP papers. However, this was done based on questions such as how well-cited papers are of different types (journal articles, conference papers, demo papers, etc.), how well-cited papers are from different areas within NLP, etc. In contrast, our work analyzes the citation pattern based on the abovementioned collaboration patterns of Academia, Industry, and Hybrid. Although the authors in \cite{mohammad2020examining} did mention that one of their future works is to measure the involvement of industry in NLP publications over time, we still haven’t seen a paper regarding it from the authors.

In terms of collaboration between industry vs academia, there have been few works in domains other than NLP. For instance, researchers in \cite{http1} discovered that between 2012 and 2016, companies wrote only 2\% of the papers in the Nature index. Most of the articles published by companies were done in cooperation with academia or governmental research institutions. Similarly, authors in \cite{http2} studied papers in the Scopus database and found out that globally, there are an increasing number of papers that were co-authored by a research institution and an industry partner between 2015 and 2019. A very recent study was done by authors in \cite{chen2022exploring}. Using a 10-year publication dataset from 2009 to 2018 retrieved from Web of Science (WoS), authors in \cite{chen2022exploring} used a bibliometric-enhanced topic modelling method to profile the main contributors of industry and academic collaboration hotspots in digital transformation research. In contrast to this paper, the authors categorized the WoS records into four types: academic authored (labelled as academic), industrial-academic co-authored (labelled as industrial), error and anonymous. All in all, the aforementioned works based on industry-academia collaboration are done either in a different domain other than NLP and/or use different categorizations compared to this paper.

\section{Methodology}

To achieve our goal of understanding who to collaborate with, we have to first analyze the collaboration patterns and citation rates between industry and academia. Hence, in this section, we will present the methodology, depicted graphically in Figure~\ref{fig:results2} to achieve the aforementioned goal. In particular, it contains three phases: 1) Data Collection, 2) Data Processing, and 3) Data Analysis.

In the data collection phase, papers from the NLP domain are retrieved and stored in the local storage as pdfs. Each pdf file is passed through the FirstPageExtracter module to only store the first page of the pdf that contains the requisite metadata, saving space as a result. Metadata includes the title, authors' affiliations, and the Digital Object Identifier (DOI) which are extracted from the TitleExtracter, AffiliationExtracter and DoiExtracter modules. Compared to retrieving the affiliations from DBLP(Computer Science Bibliography), the AffiliationExtracter module uses the pdf paper itself as it contains the most accurate affiliation when the paper was published, whereas DBLP author affiliation changes from time to time.

The data processing phase consists mainly of classifying the affiliations and finding the paper citations.

For classifying the affiliations, the Classifier module classifies the NLP papers based on each author affiliation into three distinct categories of collaboration: 1) Academia - all authors in each paper are from academia, where we define academia to include only universities and colleges. 2) Industry - all authors in each paper are from industry, where we define industry to include companies and corporations and 3) Hybrid - all authors in each paper have at least one author from academia and one from the industry. Furthermore, we count affiliations as hybrid if they belong to entities like think tanks, national labs, government agencies, or other research institutes. Although the aforementioned establishments can also be considered as part of academia, it depends on who you talk to and is not something that is cast in stone per se. There are arguments for both sides of it, and it's effectively a matter of semantics. For this paper, due to the grey area associated with this delineation and for the sake of convenience, we chose to classify them also as hybrid and not as part of academia.

For getting the paper citations, we use the DOI and the paper's title to uniquely identify any paper and pass it on to the CitationFinder module to obtain the paper's citations. The CitationFinder module extracts the Total Citations and the 3-year citations for a paper starting from its publication date. The latter is mainly computed to compare the citation impact of papers published in different years. Furthermore, it can also reveal inherent biases against famous researchers, universities, companies etc., to gain an initial boost in citations because of their status in society. To make the comparison fair, we choose to retrieve the citations obtained within three years after a paper is published. Typically, a 3-year total citation can measure the impact and quality of research work.

After we collect the necessary data and classify the articles, we will rank affiliations by paper productivity for three categories in the data analysis phase. The rank difference can show the academia and industry's attitudes towards a hybrid collaboration. Second, we will aggregate the data by year to reveal the collaboration trend. Finally, we will compare the citation distributions of the three categories to show the difference in paper impact between the three collaboration patterns.  

\begin{figure*}[htbp!]
  \centering
  \includegraphics[width=0.9\linewidth]{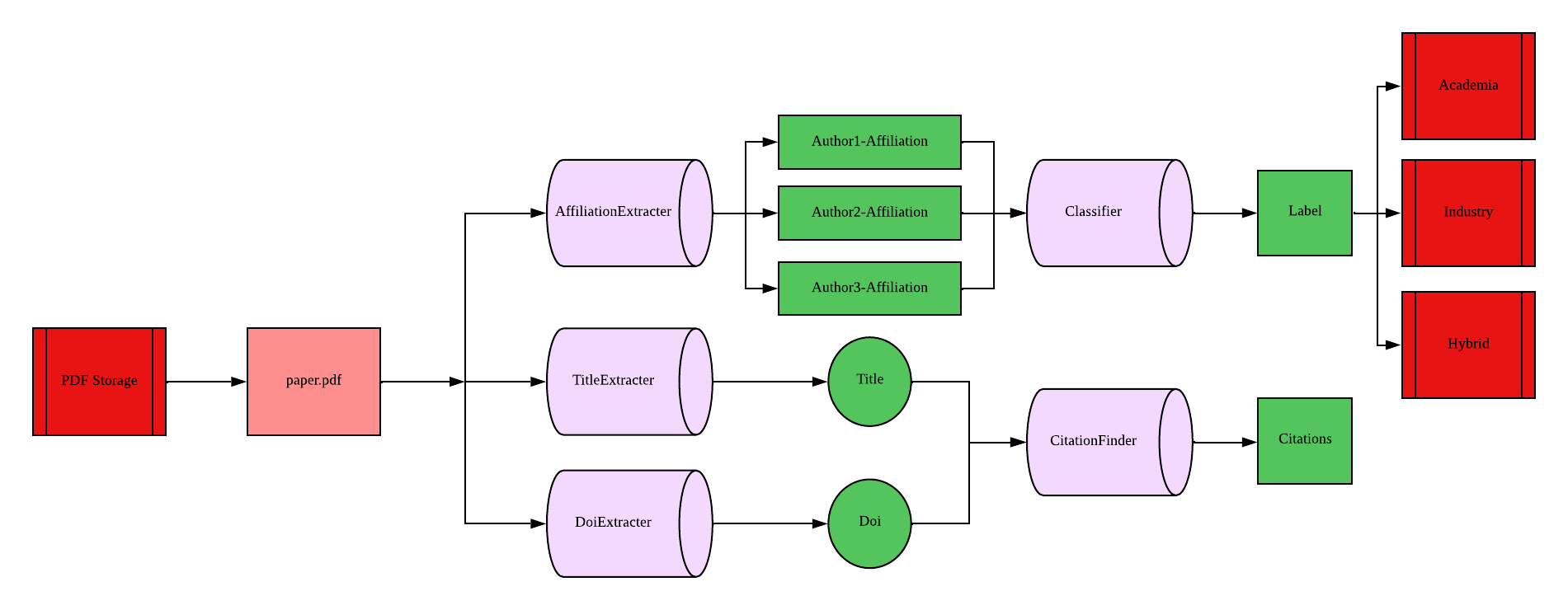}
  \caption{Pipeline}
  \label{fig:results2}
\end{figure*}

\section{Experiments}
In this section, we will present the specific implementation details of the methodology shown in the previous section. We will first show the data source and then present how we process the data to obtain the metadata. After that, we will present the data analysis and the final results. Lastly, we will give the limitations of our implementation. The code of our implementation is publicly available on GitHub. \footnote{\href{https://github.com/zbh888/research-collaboration}{\texttt{https://github.com/zbh888/research-collaboration}}}.
\subsubsection{Data Collection}
For retrieving papers in the NLP domain, we chose to use the ACL Anthology reference corpus. The ACL Anthology is a digital repository of journal and conference articles in computational linguistics and natural language processing. Its main function is to act as a reference collection for research findings. We chose to use this as it is by far the most comprehensive single repository of scientific material on NLP. In it, we chose to extract papers from the following top conferences/journals (around 13000 papers): 
\begin{itemize}
    \item Annual Meeting of the Association for Computational Linguistics (ACL)
    \item Empirical Methods in Natural Language Processing (EMNLP)
    \item Computational Linguistics (CL)
    \item North American Chapter of the Association for Computational Linguistics (NAACL)
    \item Computational Natural Language Learning (CoNLL)
    \item International Conference on Spoken Language Translation (IWSLT)
\end{itemize}
For the range of years, we chose to collect all papers published between 2012 and 2020 from the aforementioned venues. The discipline of NLP has changed significantly over the years. Still, one of the most significant changes occurred in 2012 (after the Deep Learning boom) when a few key ideas that had been developed over many years of research came to fruition. 2012 was a breakthrough year when many significant advancements in deep neural network performance and accuracy occurred \cite{krizhevsky2017imagenet}. Hence, we expect a natural proliferation of research from 2012 with respect to productivity and collaboration within individual domains and across different domains.
\subsubsection{Data Processing}
For retrieving affiliations, we used an NLP tool called “GROBID”\cite{GROBID}. It is used to extract, parse, and restructure unstructured documents like PDF into documents that are encoded in XML and TEI, with a special emphasis on technical and scientific publications. Among the multitude of functionalities it provides, we used it for header extraction to retrieve affiliations in each paper.

To achieve the aforementioned classification, we used a two-step approach. In the first step, we built two dictionaries of popular company names and university names and used keyword matching to assign each paper a classification tag of 0 or 1, denoting Academia and Industry respectively. Papers with affiliations not found in our dictionaries are assigned a classification tag of 2, denoting hybrid. In the second step, we performed a manual row-by-row scan of those papers that have a classification tag of 2 associated with them. This is because these particular sets of papers can be either academic, industry, or hybrid as their names were not in our initial dictionaries. Hence, this time-consuming process involved numerous google searches to find out about the type of institution. 

For retrieving the citations of the downloaded papers, we used Semantic Scholar (SS). It has its own academic graph to represent the world’s scientific publications. Citations of one paper connect each paper and its authors by another. In particular, we use the SS API to retrieve the three-year citations, which is a derived citation metric from the citation history data provided by SS and is mainly computed to compare the citation impact of papers published in different years.

\subsection{Data Analysis}
\begin{table}[hbt!]
\begin{center}
\caption{Top5 Affiliations}\label{A}
\begin{tabular}{||c c||} 
 \hline
 University & Company \\ [0.5ex] 
 \hline\hline
 \textcolor{blue}{Carnegie Mellon University} -4th & \textcolor{blue}{IBM} \\ 
 \hline
 University of Edinburgh - 10th & \textcolor{blue}{Microsoft}  \\
 \hline
 Stanford University -8th & \textcolor{blue}{Google}  \\
 \hline
 Johns Hopkins University -15th & \textcolor{blue}{Meta} \\
 \hline
 University of Pennsylvania -29th & Amazon \\ [1ex] 
 \hline
\end{tabular}
\end{center}
\end{table}
First, we present a high-level picture of the collaboration patterns of top affiliations. Then we will provide collaboration trends of different categories over time. Finally, we will demonstrate the differences between Academia-Academia collaboration, Industry-Industry collaboration, and Academia-Industry collaboration with respect to citations. 

For the first set of results,  Table~\ref{A} shows the top 5 universities and companies based on the collaboration in only Academia and Industry, respectively, and Table~\ref{Top5} shows the top 5 universities, research institutes, and companies in the collaboration setting of Hybrid. First, we can see that most named affiliations are big-name establishments in their own fields. Second, we notice that top industry companies showed a consistent pattern that they are equally open to collaboration within themselves (Table 1) as well as collaborating with other academia and research institutes (Table 2). This is based on the observation that the top 5 companies have similar rankings in both tables. On the other hand, academia shows a different pattern, where some universities seem to prefer collaboration more within themselves (Table 1) and others prefer collaborating with Industry and Research Institutes more (Table 2). This is based on the observation that the rankings are different for the top 5 universities in both Table 1 and Table 2.

\textbf{In summary, we can conclude that the top industry companies are both open to collaboration within themselves and with other universities and research institutes, whereas some universities prefer collaborating with other universities and some prefer more collaborations with industry and research institutes.}

\begin{table*}[hbt!]
\begin{center}
\centering
\caption{Top5 Hybrid Affiliations}\label{Top5}
% \begin{adjustbox}{max width=\linewidth}
\begin{tabular}{||c  c  c||} 
 \hline
 University & Research Institute & Company \\ [0.5ex] 
 \hline\hline
 University of the Chinese Academy of Sciences -13th & Allen Institute for AI & \textcolor{blue}{Microsoft}  \\ 
 \hline
 University of Washington -6th & German Research Centre for Artificial Intelligence & \textcolor{blue}{IBM}  \\
 \hline
 Harbin Institute of Technology -22nd  &  National Research Council &Tencent  \\
 \hline
 \textcolor{blue}{Carnegie Mellon University} -1st & RIKEN Center for Advanced Intelligence Project & \textcolor{blue}{Google} \\
 \hline
 Tsinghua University -11th & Fondazione Bruno Kessler  & \textcolor{blue}{Meta}  \\ [1ex] 
 \hline
\end{tabular}
% \end{adjustbox}
\end{center}
\end{table*}
For the second set of results, Figures 2,3 and 4 show different trends for each of the three categories of Academia, Industry and Hybrid based on different attributes.  Figure~\ref{fig:portion2} shows the trend over time based on the median of the 3-year citations for the three categories. We can observe that  both industry and hybrid papers have an overall increasing median over time compared to academia. Figure~\ref{fig:portion} and Figure~\ref{fig:growrate} show the portion and the portion growth rate over time of academia, industry and hybrid publications relative to the total accepted publications. From Figure 3, we can see that the portion of publications is increasing for both Industry and Hybrid compared to Academia which is decreasing. In terms of the growth rate, Figure 4 shows that the Academia growth rate is steady over time compared to Industry and Hybrid, which is fluctuating.

\textbf{In summary, we can conclude that the quality of industry and hybrid papers are getting better as time goes by based on median three-year citations and there is an increasing
portion of the industry and hybrid papers being published over time.}

\begin{figure}[htbp!]
  \centering
  \includegraphics[width=1.1\linewidth]{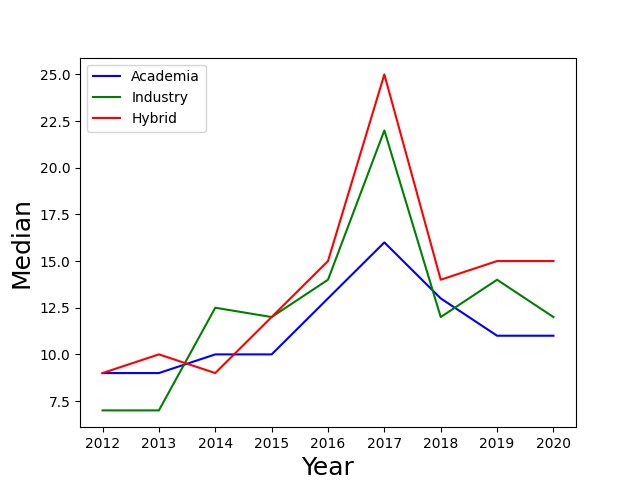}
  \caption{The Median of Three Categories}
  \label{fig:portion2}
\end{figure}

For the final set of results, we demonstrate the citation impact of papers based on 3-year citations for all three categories.

Figure~\ref{fig:CDF} shows the cumulative distribution of three classes. We can see for the same citation numbers $c$, the academia curve is majorly above both industry and hybrid categories. This means that it has a smaller portion of papers with citations higher than $c$. Table~\ref{portion} is shown to further elucidate this where the threshold represents the percentage of papers with citations greater than or equal to the given threshold. We can see that for lower citation thresholds (1 to 10), hybrid papers have the highest percentage of papers compared to the other two categories. On the other hand, for higher citation thresholds, industry papers have the highest percentage of papers.  Overall, we can see that for most citation thresholds, Academia papers have the lowest percentage of papers compared to the other two categories with the exception of the citation threshold of 2000.

\begin{figure}[htbp!]
  \centering
  \includegraphics[width=1.1\linewidth]{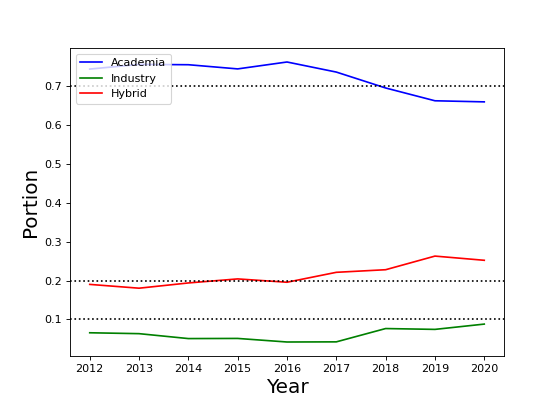}
  \caption{The Portion of Three Categories}
  \label{fig:portion}
\end{figure}

 \textbf{In summary, we can conclude that a higher percentage of Hybrid papers receive more citations within a small citation threshold, industry papers have a higher probability to produce higher-impact papers (citation threshold of 50 to 1000) even though academia still has the most impactful papers (citation threshold of 2000)}.

\begin{figure}[htbp!]
  \centering
  \includegraphics[width=1.1\linewidth]{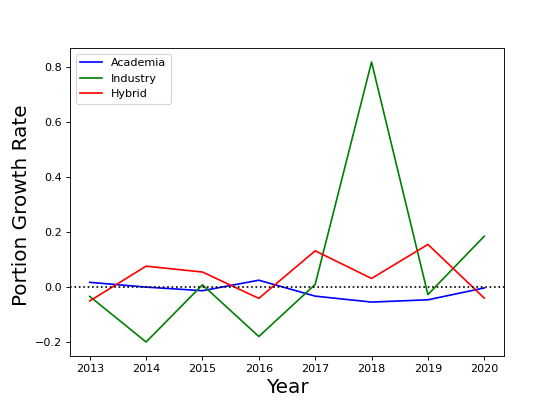}
  \caption{Portion Growth Rate Comparison}
  \label{fig:growrate}
\end{figure}

\begin{table}[h]
\begin{center}
\caption{Portion}\label{portion}
\begin{tabular}{||c c c c||} 
 \hline
 Threshold & Academia & Industry  & Hybrid \\ [0.5ex] 
 \hline\hline
$\ge$1&{97.15\%}& \textcolor{red}{96.62\%}& \textcolor{blue}{97.96\%} \\  \hline
$\ge$3&88.34\%& \textcolor{red}{86.14\%}& \textcolor{blue}{90.72\%} \\  \hline
$\ge$5&\textcolor{red}{77.07\%}& 77.70\%& \textcolor{blue}{81.52\%} \\  \hline
$\ge$10&\textcolor{red}{54.81\%}& 56.75\%& \textcolor{blue}{62.57\%}\\  \hline
$\ge$50&\textcolor{red}{9.60\%}& \textcolor{blue}{15.42\%}& 14.01\%\\  \hline
$\ge$100&\textcolor{red}{2.86\%}& \textcolor{blue}{6.86\%}& 4.12\%\\  \hline
$\ge$200&\textcolor{red}{0.81\%}& \textcolor{blue}{1.68\%}& 1.48\%\\  \hline
$\ge$500&\textcolor{red}{0.12\%}& \textcolor{blue}{0.45\%}& 0.20\%\\  \hline
$\ge$1000&0.03\%& \textcolor{blue}{0.33\%}& \textcolor{red}{0\%}\\  \hline
$\ge$2000&\textcolor{blue}{0.01\%}& \textcolor{red}{0\%}& \textcolor{red}{0\%} \\ [1ex] \hline 
\end{tabular}
\end{center}
\end{table}

\subsection{Limitations}

A few of the limitations of this work are: 1) The parsing done by the NLP tool GROBID is not always accurate. For example, there are times when it considers the address of an institution as the affiliation. Even worse, it occasionally fails to extract the affiliations altogether, and we had to manually extract affiliations for those papers; 2) The citations extracted from Semantic Scholar may not be the most complete account for all citations that a particular paper has received, compared to Google Scholar(Without an open API) which is considered to be the world’s largest source of academic information \cite{gusenbauer2019google}; 3) The classification process of categorizing papers into Academic, Industry and Hybrid are prone to errors because it involved an extensive amount of manual labelling.

\begin{figure}[htbp!]
  \centering
  \includegraphics[width=1.1\linewidth]{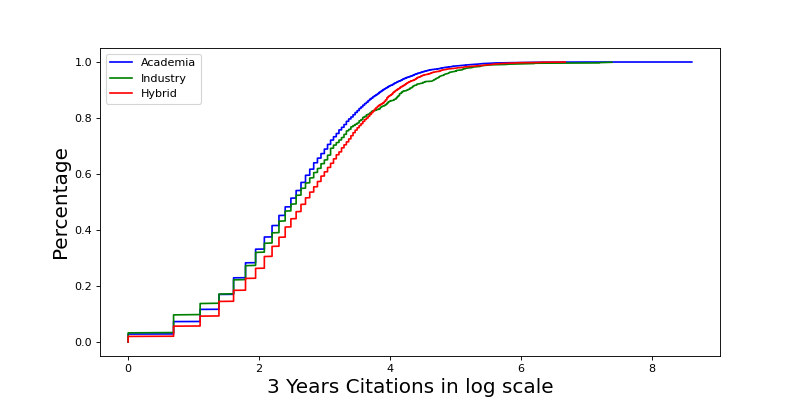}
  \caption{The Cumulative Distribution Function Plot}
  \label{fig:CDF}
\end{figure}

\section{Future Work}
% 1.Compare the Collaboration Pattern with Other Fields of Study in Computer Science
% 2. A comparitive study of collaboration pattern of different universities
% 3. Machine-learning classification of affiliations

\subsection{Compare the Collaboration Pattern with Other Fields of Study in Computer Science}
We want to extend our study to not only NLP but also other fields in computer science. By comparing our discoveries across these different fields, we can gain a more holistic understanding of the collaboration between universities and companies in the tech industry. Additionally, studying the collaboration between universities and companies in other fields will provide us with valuable insights that we can apply to our own research and potentially improve the efficiency and effectiveness of our work. 

\subsection{Comparing the Collaboration Pattern of top Companies and Universities}
We are interested in examining the collaboration patterns between leading universities and companies. It is possible that top universities may have a tendency to work with top companies due to the latter's access to ample funding. Conversely, top companies may prefer to fund research at top universities because the researchers there are more likely to be widely cited in their field. To gain further insight into this topic, we plan to identify our dataset's top universities and companies.

\subsection{Machine Learning Approach of Classification}
To classify affiliations, we employed a different approach. We first obtained a list of universities and companies from Kaggle~\cite{uniKaggle} ~\cite{compKaggle}, which provided us with a comprehensive dataset of names belonging to different institutions. We extracted only the names from the two datasets and combined them into a single list. In order to train a machine learning model to classify the names into universities and companies, we added an additional column to the dataset, labelling each name as either a university ($0$) or a company ($1$).

Next, we used the Word2Vec word embedding technique to represent each name as a numerical vector. Word2Vec is a popular method for learning word embeddings, which are low-dimensional numerical representations of words that capture the semantic and syntactic relationships between words. By representing each name as a vector, we could then apply machine learning algorithms to the dataset.

After obtaining the word embeddings for each name, we applied principal component analysis (PCA) to reduce the dimensionality of the dataset. PCA is a dimensionality reduction technique that projects high-dimensional data onto a lower-dimensional space while preserving as much of the variance as possible. By reducing the dimensionality of the data, we could make the training process more efficient and reduce the risk of overfitting.

Finally, we trained a random forest model on the resulting vectors. A random forest is an ensemble learning method that combines the predictions of multiple decision trees trained on different subsets of the data. It is a powerful and robust model that is often used for classification tasks. By training a random forest model on the word embeddings of the names in the dataset, we were able to classify them as either universities or companies with high accuracy.

We then performed a $5$-fold cross-validation to determine the optimal vector length for our word embeddings. With this approach, we were able to achieve an average accuracy of around $96\%$, which is quite promising.

However, the classification of affiliations that we extracted from the papers was not entirely accurate. This was due to the extracted affiliations being quite noisy, with many typos, abbreviations, and different languages present in the data. In addition, the formatting of the affiliations in the papers also contributed to the noise, with special characters such as "$\S$" and "$\diamondsuit$" causing confusion for our extraction function. 

We are confident in the effectiveness of this approach as long as the data has been cleaned prior to its use.

\section{Conclusion}

In this research, we developed a pipeline to extract affiliations and citations from NLP papers and used this information to create a new dataset of NLP papers. This dataset was then divided into three categories based on collaboration: academia, industry, and hybrid (academia-industry collaboration). Our analysis showed that there is a growing trend towards industry and academia-industry collaboration in NLP, and that these types of collaborations tend to produce work with a higher impact. These results suggest that academia-industry collaboration can be beneficial for researchers in academia, while researchers in the industry may already be producing high-impact work. In summary, our findings suggest that researchers in academia should consider collaborating with the industry in order to produce work with a higher impact.
%\clearpage
\bibliographystyle{IEEEtran}
\bibliography{references}

\end{document}